# Magnetoelectric oscillations in quasi-2D ferrite disk particles

M. Sigalov, E.O. Kamenetskii, and R. Shavit

Microwave Magnetic Laboratory
Department of Electrical and Computer Engineering
Ben Gurion University of the Negev, Israel
Email: kmntsk@ee.bgu.ac.il

**Abstract**

In this paper we show that magnetic-dipolar-mode (MDM) oscillations of a quasi-2D ferrite disk are characterized by unique symmetry features with topological phases resulting in appearance of the magnetoelectric (ME) properties. The entire ferrite disk can be characterized as a pair of two, electric and magnetic, coupled eigen moments. However, there is no a "glued pair" of two dipoles. An external electromagnetic field in the near-field region "views" such a ME particle, as a system with a normally oriented eigen electric moment and an in-plane rotating eigen magnetic dipole moment.

## 1. Introduction

An idea of possible existence of a local coupling between electric and magnetic dipoles in an electromagnetic medium is not new. Tellegen considered an assembly of electric-magnetic dipole twins, all of them lined up in the same fashion (either parallel or anti-parallel) [1]. Since 1948, when Tellegen suggested such "glued pairs" as structural elements for composite materials, the electrodynamics of these complex media was a subject of serious theoretical studies (see, e.g. [2 – 4]). Till now, however, the question of realization of a Tellegen medium is a subject of numerous discussions. In this paper we show that a quasi-2D ferrite disk with the MDM oscillation spectrum can be characterized as a pair of two, electric and magnetic, coupled eigen moments. But there is no a "glued pair" of two dipoles. For an external electromagnetic field such a ME particle is "viewed" as a system with a normally oriented eigen electric moment and an in-plane rotating eigen magnetic dipole moment.

## 2. The energy eigenstates, eigen electric fluxes and anapole moments in MDM ferrite disks

Generally, in classical electromagnetic problem solutions for time-varying fields, the potentials are introduced as formal quantities for a more convenient way to solve the problem and a set of equations for potentials are equivalent in all respects to the Maxwell equations for fields [5]. The situation can become completely different if one supposes to solve the electromagnetic boundary problem for wave processes in small samples of a strongly temporally dispersive medium. In small ferrite samples with strong temporal dispersion of the permeability tensor: $\vec{\vec{\mu}} = \vec{\vec{\mu}}(\omega)$, variation of the electric energy is negligibly small compared to variation of the magnetic energy and so one can neglect the electric displacement current in Maxwell equations [6]. These magnetic samples exhibit the magnetostatic (MS) resonance behaviours in microwaves, which are described by the MS-potential wave functions $\psi(\vec{r},t)$.

Such a MS-potential-wave-function description stands out also against a background of the magnetization-fluctuation description. MS ferromagnetism has a character essentially different from exchange ferromagnetism [7] and the propagating MS fields are not the exchange-interaction magnetization waves.

For MDMs in a quasi-2D ferrite disk described by MS-potential wave functions $\psi(\vec{r},t)$ one has evident quantum-like attributes and the spectrum is characterized by energy eigenstate [8, 9]. It was shown, however, that because of the boundary condition on a lateral surface of a ferrite disk, membrane MS functions cannot be considered as single-valued functions. This leads to the topological effects which result in appearance of the fluxes of pseudo-electric fields and eigen electric moments – the anapole moments [9]. An observation of such anapole moments was realized in recent microwave experiments [10].

The complete-set MDM oscillating spectrum is obtained in neglect of the electric displacement currents and so there should be no influence of a dielectric loading on the spectral peak positions in a ferrite disk. Nevertheless, our experiments show such an influence. In experiments we used a ferrite sample of a diameter $2\Re = 3\,\text{mm}$ made of the YIG film on the GGG substrate (the YIG film thickness $d = 50\,\text{mkm}$, saturation magnetization $4\pi M_0 = 1880\,\text{G}$, linewidth $\Delta H = 0.8\,\text{Oe}$) and a short-wall rectangular waveguide. A normally magnetized ferrite disk was placed in a maximal RF electric field of the TE10 mode and is oriented normally to the *E*-field. We analyze the MDM spectra with respect to frequency at constant DC magnetic field. A bias magnetic field is $H_0 = 4900\,\text{Oe}$. A ferrite disk is placed on a GGG substrate which has the dielectric permittivity parameter of $\varepsilon_r = 15$. Now we put dielectric samples above a ferrite disk. There are dielectric disks of a diameter 3 mm and thickness 2 mm. We used a set of disks of commercial microwave dielectric (non magnetic) materials with the dielectric permittivity parameters of $\varepsilon_r = 15$ (K-15; TCI Ceramics Inc), $\varepsilon_r = 30$ (K-30; TCI Ceramics Inc), and $\varepsilon_r = 100$ (K-100; TCI Ceramics Inc). We observed strong variations of the spectral pictures when dielectric disks were placed above a ferrite disk. These transformations of the spectra become most evident when we match (by proper small shifts of the bias magnetic fields) positions of the first peaks in the spectra. From the absorption spectra shown in Fig. 1 one can see that as the dielectric permittivity parameter of a dielectric sample increases, the frequency shift of the mode peak position increases as well.

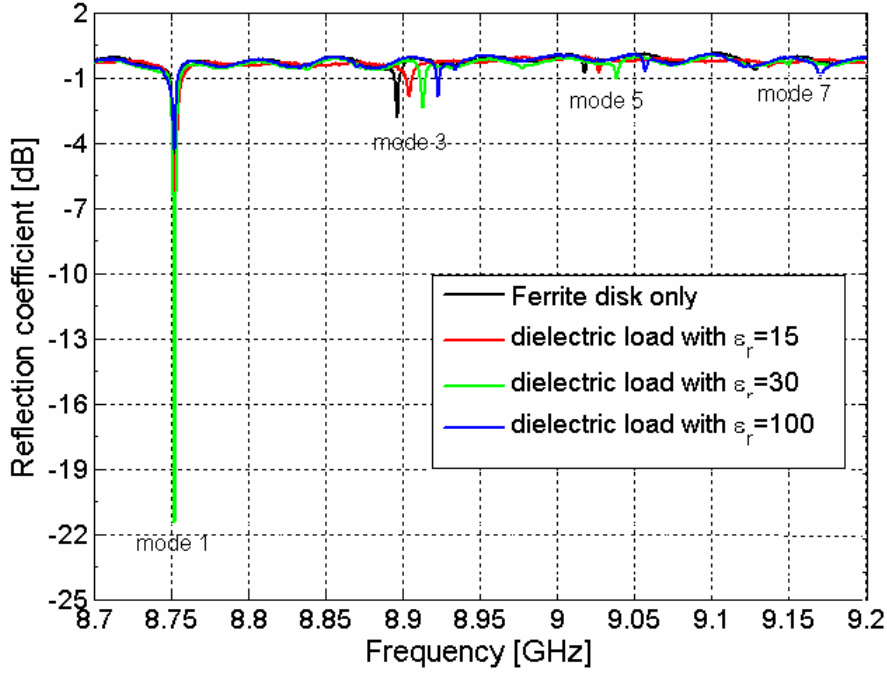

Fig. 1: The frequency shift of the odd MDM peak positions for a ferrite disk with dielectric loading due to the eigen electric fluxes. The first peaks in the spectra are matched by proper correlations of bias magnetic fields.

**3. In-plane rotating eigen magnetic dipole moments**

From the solved spectral problem for MS-potential wave functions, one can obtain the field distributions inside a quasi-2D ferrite disk. It can be easily shown that for the eigen curl electric field inside a disk one has the Poisson equation:

$$\nabla^2 \vec{E} = 4\pi \vec{j}^{\,e}, \qquad (1)$$

where $\vec{j}^{\,e} \equiv i\omega \vec{\nabla} \times \vec{m}$, $\vec{m}$ is RF magnetization. For a normally magnetized quasi-2D disk, the eigen curl electric fields can be experimentally observed via appearance of effective magnetic dipoles. The electric field has the in-plane orientation with the contrarily directed vectors on the upper and lower disk planes. Recently, it was shown that eigenfunction patterns of a numerically solved (based on the HFSS program) non-integrable electromagnetic problem of a quasi-2D ferrite disks in a rectangular-waveguide cavity are in a very good correspondence with the analytically studied MDMs [11]. This allows making a detailed analysis of the field patterns inside a ferrite disk. Fig. 2 gives a top view of the numerically solved electric field distributions for the first MDM on the upper plane of a ferrite disk at different time phases.

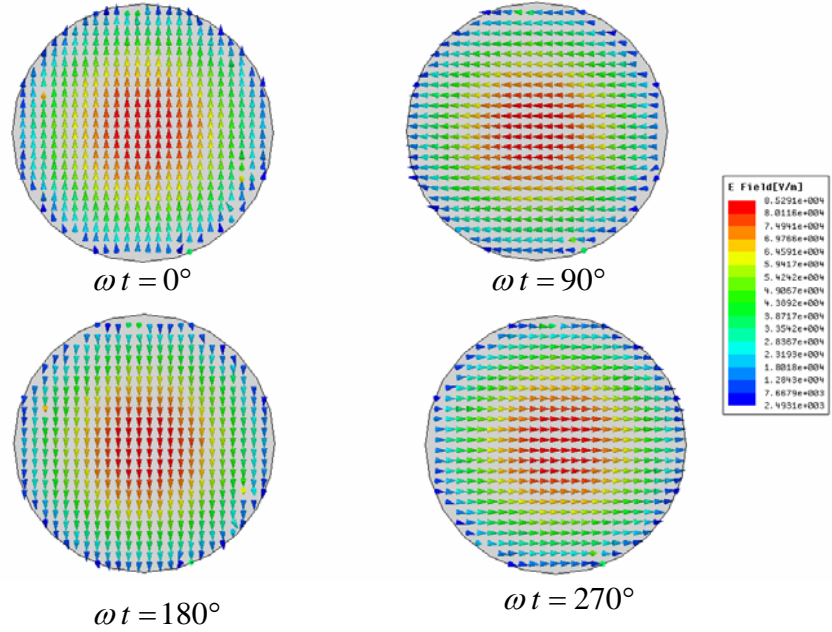

$\omega t = 0°$ $\omega t = 90°$

$\omega t = 180°$ $\omega t = 270°$

Fig. 2: A top view of electric field distributions for the first MDM on the upper plane of a ferrite disk at different time phases. In-plane linear magnetic currents are oriented perpendicular to the electric field vectors.

The in-plane electric fields on the upper and lower planes of a ferrite disk are in opposite directions at any time phase. Since the disk thickness is much, much less than the rectangular waveguide height, the disk in a cavity can be clearly replaced by a magnetic-current sheet. A surface density of the effective magnetic current is expressed as

$$\vec{n} \times (\vec{E}_{upper} - \vec{E}_{lower}) = -\frac{4\pi}{c} \vec{i}^{\,m}, \qquad (2)$$

where $\vec{E}_{upper}$ and $\vec{E}_{lower}$ are, respectively, in-plane electric fields on the upper and lower planes of a ferrite disk and $\vec{n}$ is a normal to a disk plane directed along a bias magnetic field. Evidently, $\vec{E}_{upper} = -\vec{E}_{lower}$. Following pictures in Fig. 2, one can conclude that there are rotating linear magnetic currents. Non-zero current-line divergence of such magnetic currents gives equivalent magnetic charges. As a result of this equivalent representation, one has an evidence for an in-plane rotating magnetic dipoles for the entire ferrite disk with MDM oscillations. Due to such rotating magnetic dipoles one has excitation of MDMs in a quasi-2D ferrite disk shown in well-known experiments [12].

## 4. Conclusion

Based on experimental studies and numerical simulation results we showed that a quasi-2D ferrite disk with MDM oscillations can be characterized as a system of a normally oriented eigen electric moment and an in-plane rotating eigen magnetic dipole moment. The observed characteristics of ferrite-based microwave ME particles arise from the fact that MDM oscillations in quasi-2D ferrite disks are macroscopic quantum coherence states with discrete energy levels and topological vortex structures.